\documentclass[english]{article}
\usepackage{lmodern}
\usepackage[T1]{fontenc}
\usepackage[latin9]{inputenc}
\usepackage{amsmath}
\usepackage{amssymb}
\usepackage{graphicx}
\usepackage{babel}
\usepackage{geometry}
\usepackage{authblk}

\begin{document}

\title{A cosmic accelerated scenario based on degrees of freedom of the spacetime}

\author[a]{Fei-Quan Tu \thanks{Corresponding author: fqtuzju@foxmail.com}}
\author[b]{Yi-Xin Chen}
\author[a]{Qi-Hong Huang}
\affil[a]{School of Physics and Electronic Science, Zunyi Normal University, Zunyi 563006, China}
\affil[b]{Zhejiang Institute of Modern Physics, Zhejiang University, Hangzhou 310027, China}


\renewcommand*{\Affilfont}{\small\it} 
\renewcommand\Authands{ and } 
\date{} 
\maketitle

\begin{abstract}

Several recent investigations have shown that there is a holographic relationship between the bulk degrees of freedom and the surface degrees of freedom in the spacetime. Furthermore, the entropy on the horizon can produce an entropic force effect on the bulk degrees of freedom. In this paper, we explore the dynamic evolution law of the universe based on the idea of the entropic force and asymptotically holographic equipartition and further analyze the thermodynamic properties of the current model. We get the age of the universe, the relation between the luminosity distance and the redshift factor and the deceleration parameter which are consistent with astronomical observations. In addition, we can well explain the age of the universe and the mechanism of accelerated expansion without introducing dark energy for the evolution history of the universe up to now. We also show that the generalized second law of thermodynamics, the energy balance condition and the energy equipartition relation always hold. More importantly, the energy balance condition is indeed a holographic relation between the bulk degrees of freedom and the surface degrees of freedom of the spacetime. Finally, we analyze the energy conditions and show that the strong energy condition is always violated and the weak energy condition is satisfied when $t\leq2t_{0}$ in which $t$ is the time parameter and $t_{0}$ is the age of the universe.

\end{abstract}

\section*{1. Introduction }

It has been found that the late universe is in accelerated expansion
according to astronomical observations \cite{key-1,key-2} since the
1990s. To resolve the problem of the accelerated expansion of the universe, various
cosmological models have been proposed. One of the most common models is to introduce
a cosmological constant, i.e., the state parameter $\omega$
is $-1$ for the equation of state $p=\omega\rho$. Also, recent observations
show that the parameter $\omega$ for the dark energy could be dynamical,
i.e., the state parameter can be expressed as $\omega(t)$ \cite{key-3,key-4}.
In addition, the thermodynamic scenarios have been used to investigate the evolution of the universe widely.

In the past few decades, it has been pointed out that the action can be
decomposed into the bulk term and surface term in a wide class of
gravitational theories \cite{key-5,key-6}. We often assume a boundary
condition or add a term to cancel the surface term to derive the gravitational
field equation when we apply the variational principle to the action.
Thus the field equation derived by this method is only related
to the properties of the bulk term of the action and has nothing
to do with the properties of the surface term. However, it has been shown
that there is a key relationship between the surface term and
the bulk term \cite{key-7}. Besides, the equation of gravitational
field can be derived from the surface term in the homeomorphism
invariant gravity \cite{key-8} and the entropy
of the horizon can be given when one evaluates the surface term on the horizon \cite{key-9,key-10}.
More importantly, it has been shown that there exists a direct connection between
the surface term in the action principle and the thermodynamics of the null surface \cite{key-11}.
These scenarios describing cosmology are based on the horizon thermodynamics
which is closely related to holographic principle \cite{key-12,key-13}:
all information in the three-dimensional space surrounded by a surface
can be described by the two-dimensional surface. Among the scenarios,
the entropic force model \cite{key-14,key-15,key-16} and Padmanabhan's holographic equipartition
model \cite{key-17,key-18} are introduced to study the accelerated expansion of the universe.

In the entropic force model, an additional term is phenomenally introduced
to the accelerated equation of the FRW universe due to the existence
of the surface term of the action. Based on the entropic effect of
the Hubble horizon which is equivalent to creating a negative pressure
on the cosmic matter, the current accelerated expansion \cite{key-15}
and the inflation at the early period \cite{key-16} can be described
well. This model is widely used to study the evolution of the universe
\cite{key-19,key-20,key-21,key-22,key-23,key-24,key-25,key-26,key-27}.
In the holographic equipartition model, the dynamic evolution of the universe can be described
by the difference between the number of degrees of freedom on the
surface and in the bulk \cite{key-17,key-18}.
By calculating the surface degrees of freedom in different entropy
forms (e.g., Tsallis entropy, Renyi entropy or quantum correction
entropy), one can get the different dynamic evolution equation with
an extra driving term \cite{key-28,key-29,key-30,key-31,key-32,key-33,key-34,key-35,key-36}.
This driving term corresponds to a time-varying ``cosmological constant''.
Thus the cosmological issues can be discussed from the thermodynamic
viewpoint. These approaches give a new insight into the problem of quantum gravity.

Since the surface degrees of freedom exist on the cosmological horizon
which produces the entropic force on matter in the bulk, we should
consider the entropic force effect for the dynamic evolution of the
universe. In this paper, the entropic force effect is considered as
an extra negative pressure on matter in the bulk. Under this consideration,
we modify the number of degrees of freedom in the bulk and thus the
dynamic evolution equation of the universe. Furthermore, we consider
a time-varying equation of state by combining the equation of state
in the photon dominated period with that in the pressureless
matter dominated period. In this way, we get the age of the universe,
the relation between the luminosity distance and the redshift factor
and the deceleration parameter which are consistent with astronomical observations.
Next, we analyze the thermodynamic properties of the current
model and find that the generalized second law of thermodynamics,
the energy balance condition and the energy equipartition relation always hold.
Finally, we analyze the energy conditions and show
that the strong energy condition is always violated and the
weak energy condition is satisfied when $t\leq2t_{0}$.

The paper is organized as follows. In Section 2, we derive the dynamic
evolution equation of the universe based on the idea of the entropic
force and asymptotically holographic equipartition. Combined with the
equation of state of matter changing with time, we obtain the evolution
solutions of the universe and analyze the physical meaning of these
solutions. In Section 3, we analyze that the generalized second law
of thermodynamics, the energy balance relation, the energy equipartition relation
and the strong and weak energy conditions.
Finally, we present the conclusions and discuss
the fluid viscosity and particle creation models. Throughout this
paper we take the convention $k=c=\hbar=1$ for the sake of simplicity.

\section*{2. Dynamical evolution of the universe caused by the entropic force and asymptotically
holographic equipartition}

For the gravitational Lagrangian $L_{g}$ , it can be expressed as
\begin{equation}
16\pi\sqrt{-g}L_{g}=L_{bulk}+L_{sur},
\end{equation}
where $L_{bulk}$ is the bulk term and $L_{sur}$ is the surface term
related by the equation $\sqrt{-g}L_{sur}=\partial_{\lambda}\left(g_{\mu\nu}\frac{\partial(\sqrt{-g}L_{bulk})}{\partial(\partial_{\lambda}g_{\mu\nu})}\right)$
\cite{key-9,key-10,key-37}. This relation is usually called holographic
relation and allows the surface term to contain information about
the bulk. So if we take the action related to this gravitational Lagrangian
$L_{g}$ as the underlying starting point in the gravitational theory,
then we have to consider the holographic effect of the surface term.

Moreover, the entropy is given when we evaluate
the surface term on the horizon \cite{key-9,key-10}. Hence,
we can consider the holographic effect as the effect caused by the
horizon entropy. Here we employ the Hubble horizon as the cosmological
horizon, i.e., the radius of the horizon $R_{H}=1/H$. According to
Bekenstein's result \cite{key-38,key-39}, the horizon entropy is
\begin{equation}
S_{h}=\frac{A}{4L_{p}^{2}}=\frac{\pi R_{H}^{2}}{L_{p}^{2}}=\frac{\pi}{H^{2}L_{p}^{2}},
\end{equation}
where $A$ is the area of the horizon and $L_{p}^{2}$ is the Planck
area. The variation of energy with respect to the radius $r$ gives
the entropic force \cite{key-15,key-16,key-19,key-20}
\begin{equation}
F=-T\frac{dS_{h}}{dr}=-\frac{1}{L_{p}^{2}},
\end{equation}
where we choose the Hubble horizon $r=R_{H}$. The entropic force leads to an extra
pressure
\begin{equation}
p_{e}=\frac{F}{A}=-\frac{H^{2}}{4\pi L_{p}^{2}}.
\end{equation}
From this result, we can see that the surface term of the gravitational
Lagrangian $L_{g}$ has the negative pressure effect similar to that
of dark energy or cosmological constant.

In order to explain the accelerated expansion of the universe, the holographic equipartition
model based on gravitational thermodynamics is proposed \cite{key-17,key-18}.
In such model, the difference between the surface
degrees of freedom, $N_{sur}$, and the bulk degrees of freedom, $N_{bulk}$,
in a region of space drives the accelerated expansion of the universe
through a simple equation
\begin{equation}
\frac{dV}{dt}=\left(N_{sur}-N_{bulk}\right)L_{p}^{2},
\end{equation}
where $V=4\pi/3H^{3}$ is the Hubble volume and $t$ is the cosmic time.
The number of surface degrees of freedom can be expressed as \cite{key-30,key-31}
\begin{equation}
N_{sur}=4S_{h}=\frac{4\pi}{H^{2}L_{p}^{2}},
\end{equation}
while the number of bulk degrees of freedom is assumed to satisfy
the equipartition law of energy

\begin{equation}
N_{bulk}=\frac{|E_{K}|}{\frac{1}{2}T},
\end{equation}
where $|E_{K}|$ is taken as the Komar energy $|(\rho+3p)|V$ contained
inside the Hubble volume $V=(4\pi/3H^{3})$ and $T=\frac{H}{2\pi}$
is the temperature of the bulk degrees of freedom.
Considering the entropic force effect on the horizon, we take the Komar
energy as
\begin{equation}
|E_{K}|=|\rho+3(p+p_{e})|V,
\end{equation}
where $p_{e}$ is given by Eq.(4), $\rho$ and $p$ are the density
and pressure of matter respectively. Thus we obtain the number of
bulk degrees of freedom
\begin{equation}
N_{bulk}=-\frac{(4\pi)^{2}}{3H^{4}}(\rho+3p)+\frac{4\pi}{H^{2}L_{p}^{2}}.
\end{equation}
Substituting the Hubble volume, Eq.(6) and Eq.(9) in Eq.(5), we
obtain the evolution equation of the universe
\begin{equation}
\frac{\ddot{a}}{a}=H^{2}+\dot{H}=-\frac{4\pi L_{p}^{2}}{3}(\rho+3p)+H^{2}.
\end{equation}
Compared with the standard cosmic acceleration equation of the FRW universe
(i.e., the cosmological constant is not taken into account), we find
that there is an additional term $H^{2}$ which leads to the accelerated
expansion of the universe. In the lambda cold dark matter $(\mathrm{\Lambda CMD})$
model, the cosmic acceleration equation of the FRW universe is $\frac{\ddot{a}}{a}=-\frac{4\pi L_{p}^{2}}{3}(\rho+3p)+\frac{\Lambda}{3}$,
where $\frac{\Lambda}{3}\approx1.3\times10^{-52}m^{-2}$ \cite{key-40}.
In our model, we obtain $H_{0}^{2}\approx0.6\times10^{-52}m^{2}$
by substituting the value of current Hubble constant $H_{0}$. Hence, our model
is compatible with current cosmological observations in terms of the
numerical magnitude. However, we should also note that the term $H^{2}$
changes with time while the term $\frac{\Lambda}{3}$ is a constant,
thus we obtain the different evolution law from the $\mathrm{\Lambda CMD}$
model.

The continuity equation for these degrees of freedom of the spacetime
is read as
\begin{equation}
\dot{\rho}+3H(\rho+p+p_{e})=0.
\end{equation}
Here we consider the negative pressure caused by the entropic force
as an effective pressure on matter, i.e., the pressure on matter is
described by $p+p_{e}$. For the current model, the first law of thermodynamics
that the universe follows is
$d(\rho V_{c})=-(p+p_{e})dV_{c}$ where $V_{c}$ is the volume of the universe.
The evolution of $V_{c}$ can be expressed as $V_{c}=V_{0}a^{3}$ where $V_{0}$
is a positive constant. Hence Eq.(11) is derived from the first law of thermodynamics.

In order to get the properties of the dynamic
evolution of the universe, we must also consider the equation of state
of matter. Assume that the equation of state for time $t<4t_{0}$
has the following form
\begin{equation}
p=\left(-\frac{t}{3t_{0}}+\frac{1}{3}\right)\rho,
\end{equation}
where $t_{0}$ is the present time, i.e., the current age of the
universe. There are several reasons to employ this assumption. First,
in the early universe dominated by photons, the equation of state
of cosmic matter is $p=\frac{1}{3}\rho$. In contrast, the equation
of state is $p=0$ when the pressureless matter dominates the universe.
Second, the time interval from the big bang
to the photon dominated period is very short compared with the age
of the universe so that we can ignore it. Therefore, we can consider
that the universe dominated by photons evolves into the universe dominated
by the pressureless matter during the evolution history of the universe so far.
Based on these facts, we assume that the photon dominated period changes
linearly to the matter dominated period, i.e., the equation of state
changes linearly from $p=\frac{1}{3}\rho$ at $t=0$ to $p=0$ at
$t=t_{0}$. In this way, we introduce the equation of state of cosmic
matter Eq.(12) to describe the whole evolution history of the universe.
As for why we assume the condition $t<4t_{0}$, we will explain it
later.

Combining Eq.(10), Eq.(11) and Eq.(12), we obtain the solutions for
$t<4t_{0}$
\begin{equation}
\dot{H}=-\left(1-\frac{t}{2t_{0}}\right)H^{2}
\end{equation}
and
\begin{equation}
H^{2}=\frac{8\pi L_{p}^{2}}{3}\rho.
\end{equation}
From Eq.(13), we obtain
\begin{equation}
\frac{1}{H}=t-\frac{t^{2}}{4t_{0}}.
\end{equation}
Thus the age of the universe is $t_{0}=\frac{4}{3H_{0}}$.
If the value of the current Hubble constant is taken as $H_{0}=70\; km\, s^{-1}Mpc^{-1}$
\cite{key-41}, one can obtain the age of the universe
\begin{equation}
t_{0}=13.1\times10^{9}\: year,
\end{equation}
which is consistent with the data of astronomical observations $t_{0}=13.4_{-1.0}^{+1.3}\times10^{9}$
year \cite{key-41}. Hence, we obtain the age of the universe
without introducing dark energy.

Integrating Eq.(15), we obtain the scale factor
\begin{equation}
a(t)=\frac{3a_{0}t}{4t_{0}-t}
\end{equation}
and the Hubble parameter
\begin{equation}
H(t)=\frac{4t_{0}}{(4t_{0}-t)t},
\end{equation}
where $a_{0}=a(t_{0})$ is the current scale factor.

Now we investigate the relation between the luminosity distance
and the redshift factor. The general formula is given by Refs.\cite{key-25,key-42}
\begin{equation}
d_{L}=\frac{1+z}{H_{0}}\int_{1}^{1+z}\frac{dy}{H/H_{0}},
\end{equation}
where $z$ is the redshift factor defined by $z+1\equiv y=a_{0}/a$.
After substituting Eq.(17) and Eq.(18) into Eq.(19) and some simple calculations,
we obtain the relation between the luminosity distance and the redshift factor
\begin{equation}
H_{0}d_{L}(z)=\frac{16}{9}(1+z)\left[\ln\frac{4+3z}{4}+\frac{1}{4+3z}-\frac{1}{4}\right]
\end{equation}

For the $\Lambda CDM$ model, the relation between the luminosity distance and
the redshift factor in the flat universe can be expressed as \cite{key-43}
\begin{equation}
H_{0}d_{L}=(1+z)\int_{0}^{z}dz^{'}[(1+z^{'})^{2}(1+\Omega_{m}z^{'})-z^{'}(2+z^{'})\Omega_{\Lambda}]^{-1/2},
\end{equation}
where $\Omega_{m}$ is the ratio of the density of dark matter to the critical density and
$\Omega_{\Lambda}$ is the ratio of the density of dark energy to the critical density.
It has been found that the description of the universe for $\Omega_{m}=0.27$ and $\Omega_{\Lambda}=0.73$
is consistent with the supernova data in Ref.\cite{key-44}. The curve relations between the luminosity distance and the redshift factor
for the $\Lambda CDM$ model and the current model are shown in Figure 1. Our result is well in line with the result of $\Lambda CDM$ model.

\begin{figure}
\centering
\includegraphics[width=9cm]{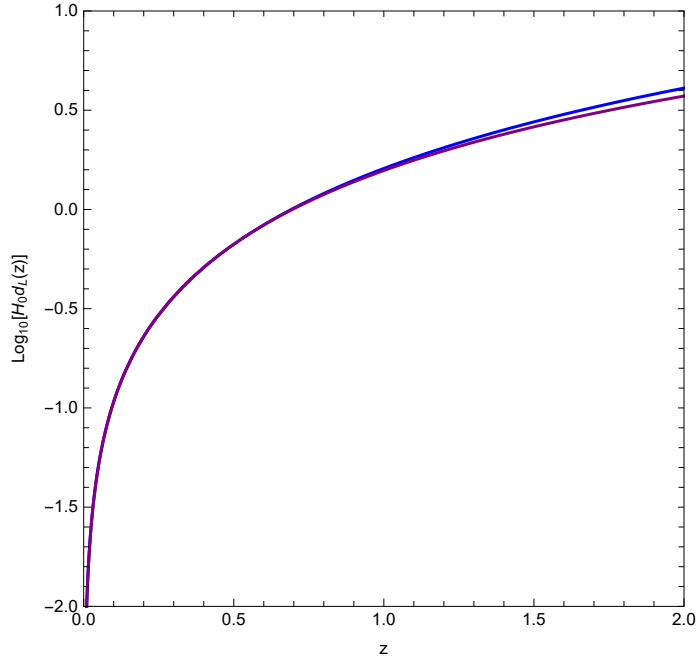}
\caption{(color online). The upper curve represents
the relation between the luminosity distance and the redshift factor
in our model, the lower curve represents the relation in $\Lambda CDM$ model.
Our result is well in line with that of $\Lambda CDM$ model.}
\end{figure}

The deceleration parameter of the universe can be expressed as
\begin{equation}
q\equiv - \frac{\ddot{a}a}{\dot{a}^{2}}=-\frac{t}{2t_{0}}.
\end{equation}
This result shows that the universe is in accelerating. At the present
time $t=t_{0}$ , we obtain $q=-\frac{1}{2}$ which is consistent
with supernova observations \cite{key-45}. From Eq.(14), we obtain
also the density of the universe
\begin{equation}
\rho=\frac{6t_{0}^{2}}{\pi L_{p}^{2}(4t_{0}-t)^{2}t^{2}},
\end{equation}
which also shows that the density of the current universe is $\rho_{0}=2/(3\pi L_{p}^{2}t_{0}^{2})$
which is of the same order of magnitude as the critical density.

From Eq.(17), we can see that there exists a singularity when $t=4t_{0}$ and
the scale factor $a(t)$ is less than $0$ when $t>4t_{0}$. Since $r=a(t)r_{0}$
where $r$ is the physical distance and $r_{0}$ is the comoving distance, the scale factor $a(t)$
has to be a finite positive value. This is why we choose the condition $t<4t_{0}$.

It should be noted that we have obtained the age, the relation between the luminosity distance and the redshift factor,
the deceleration parameter and the energy density of the universe
which are consistent with astronomical observations without introducing dark energy.
That is, our model can well explain some important cosmological observational facts
when $t\leq t_{0}$. However, our model does not exclude
the existence of dark energy because the pressure of matter is negative when
$t>t_{0}$, i.e., matter shows the nature of ``dark energy''
when $t>t_{0}$. Meanwhile, we can see from Eq.(18) and Eq.(20) that
the Hubble parameter $H(t)$ and the density of the universe $\rho(t)$
reach the minimum when $t=2t_{0}$, and then diverge when $t=4t_{0}$.
Thus the evolution of the universe
can be described as follows. The universe originates from a singularity
with extremely high density, then the horizon expands to the maximum
when $t=2t_{0}$ due to the entropic force effect and shrinks to the
initial state when $t\rightarrow4t_{0}$. In addition, we also see
from Eq.(12) that the equation of state of matter is $p=-\rho$ when $t\rightarrow4t_{0}$
which is the equation of state for the vacuum energy. Since
the scale factor grows more rapidly than the horizon, galaxies
disappear beyond the horizon and the universe becomes increasingly dark
as time progresses when $t>2t_{0}$.
There exists a Big Rip singularity when $t=4t_{0}$ \cite{key-46,key-47}.
We can also see that the remaining life span of the universe
is three times its current age for the current model.

\section*{3. Analyses of thermodynamic properties and thermodynamic constraints
for the current model}

A thermodynamic analysis for a cosmological model can lead to a better
understanding of the evolution of the universe. More importantly,
the thermodynamic constraints on the model can narrow the range of
parameters so that we can get a more reliable cosmological model.

First, let us analyze the entropy change of the current model to see
if the generalized second law of thermodynamics is satisfied, and
then give the constraints of time $t$. Here we employ the local equilibrium
hypothesis, i.e., the matter temperature equals the horizon temperature,
thus the first law of thermodynamics can be written as \cite{key-48,key-49}
\begin{equation}
TdS_{m}=dE_{m}+(p+p_{e})dV,
\end{equation}
where $S_{m}$ is the entropy of matter, $V$ is the Hubble volume
and $E_{m}$ is the Misner-Sharp energy which is given by
\begin{equation}
E_{m}=\int T_{\mu\nu}u^{\mu}u^{\nu}dV=\rho V,
\end{equation}
in which
\begin{equation}
T_{\mu\nu}=(\rho+p+p_{e})u_{\mu}u_{\nu}+(p+p_{e})g_{\mu\nu}
\end{equation}
is the energy-momentum tensor of the fluid and $u^{\mu}=\delta_{t}^{\mu}$
is a 4-velocity of a comoving observer.

Substituting Eq.(25) in Eq.(24), we obtain the entropy change of
matter inside the horizon
\begin{equation}
\dot{S}_{m}=\frac{1}{T}[(\rho+p+p_{e})\dot{V}+V\dot{\rho}].
\end{equation}
Using Eq.(11) and Eq.(14), we obtain
\begin{equation}
\rho+p+p_{e}=-\frac{\dot{H}}{4\pi L_{p}^{2}}
\end{equation}
and
\begin{equation}
\dot{\rho}=\frac{3H\dot{H}}{4\pi L_{p}^{2}}.
\end{equation}
Then substituting Eq.(28), Eq.(29), the volume of Hubble horizon $V=\frac{4\pi}{3H^{3}}$
and the temperature $T=\frac{H}{2\pi}$ in Eq.(27), we obtain the
entropy change of matter
\begin{equation}
\dot{S}_{m}=\frac{2\pi\dot{H}}{H^{3}L_{p}^{2}}\left(1+\frac{\dot{H}}{H^{2}}\right).
\end{equation}
 For the Hubble horizon, its entropy change is
\begin{equation}
\dot{S}_{h}=\frac{-2\pi\dot{H}}{H^{3}L_{p}^{2}}.
\end{equation}
So the total entropy change is
\begin{equation}
\dot{S}=\dot{S}_{m}+\dot{S}_{h}=\frac{2\pi}{HL_{p}^{2}}\left(\frac{\dot{H}}{H^{2}}\right)^{2}.
\end{equation}
From Eq.(32), we see that the total entropy change is always non-negative,
so the generalized second law of thermodynamics always holds
and the condition $t<4t_{0}$ is also reasonable in our model.

Inserting the energy density and Hubble volume into Eq.(25), we get
the Misner-Sharp energy
\begin{equation}
E_{m}=\rho V=\frac{1}{2HL_{p}^{2}}.
\end{equation}
On the other hand, we also find
\begin{equation}
TS_{h}=\frac{1}{2HL_{p}^{2}}.
\end{equation}
Thus we obtain the energy balance relation
\begin{equation}
\rho V=TS_{h},
\end{equation}
where the term $TS_{h}$ in the right hand side is the heat energy
of the Hubble horizon \cite{key-50,key-51}. This relation shows that
the Misner-Sharp energy within the Hubble volume is equal to the heat
energy on the Hubble surface. Furthermore, the surface term of the
action is found to be the integral of the heat density $Ts$ where
$s$ is the entropy density \cite{key-11}. For the Hubble horizon with
a uniform temperature distribution, $TS_{h}$ is the surface term
of the action. In addition, it has been shown that $TS_{h}$ is also
the Noether energy (or we call it Noether charge) \cite{key-52} and
can be thought as the heat content of the horizon \cite{key-53}.
Therefore, Eq.(35) also shows that there exists a holographic relationship
between the bulk degrees of freedom and the surface degrees of freedom
of the spacetime.

In the current model, the energy that drives the accelerated expansion
of the universe is the energy which corresponds to the surface degrees
of freedom, we call it the equipartition energy. The equipartition
energy is
\begin{equation}
E=\frac{1}{2}N_{sur}T=\frac{1}{HL_{p}^{2}},
\end{equation}
where we use Eq.(6) and $T=H/2\pi$. Hence we can obtain the relation
\begin{equation}
E=2TS_{h},
\end{equation}
which is consistent with the conclusion obtained by Refs.\cite{key-52,key-54}

At the end of this section, we will discuss the energy conditions.
As we know, the energy-momentum tensor of matter determines the structure
of the spacetime, but there are many kinds of matter in the spacetime and
further their equations of state are also changeable, so it is unrealistic
to find a universal energy-momentum tensor to describe all known distribution
of matter. In order to study the properties of singularities or the
form and distribution of matter, we need to find some universal conditions,
which are energy conditions \cite{key-55}. There are many kinds of
energy conditions in the literatures. Here we only discuss the strong energy condition
and the weak energy condition.

The strong energy condition states that
\begin{equation}
\left(T_{\mu\nu}-\frac{1}{2}g_{\mu\nu}T\right)u^{\mu}u^{\nu}\geq0
\end{equation}
for any timelike vector $u^{\mu}$. For the fluid whose energy-momentum
tensor is Eq.(26), the strong energy condition states
\begin{equation}
\rho+3(p+p_{e})\geq0\quad and\quad\rho+p+p_{e}\geq0,
\end{equation}
while the weak energy condition is
\begin{equation}
T_{\mu\nu}u^{\mu}u^{\nu}\geq0
\end{equation}
for any timelike vector $u^{\mu}$ which can be expressed as
\begin{equation}
\rho+p+p_{e}\geq0\quad and\quad\rho\geq0.
\end{equation}

For the current model, we obtain
\begin{equation}
\rho+3(p+p_{e})=-\frac{t}{t_{0}}\rho
\end{equation}
and
\begin{equation}
\rho+p+p_{e}=\frac{2}{3}\left(1-\frac{t}{2t_{0}}\right)\rho.
\end{equation}
From the above results, we can see that the strong energy condition
is always violated while the weak energy condition is satisfied when
$t\leq2t_{0}$. As we can see from Eq.(22), the expansion of the universe is in accelerating,
so it is a natural result that the strong energy condition
has to be violated. For the scope of time $2t_{0}<t<4t_{0}$, we cannot rule it out
although the weak energy condition is violated, because the phantom
fluid which violates the weak energy condition may also serve as another
candidate for dark energy \cite{key-56,key-57,key-58}. We also confirm that
the continuity equation (11) based on the first law of thermodynamics
is equivalent to the conservation condition $\nabla^{\mu}T_{\mu\nu}=0$.

\section*{4. Conclusions and discussions}

In this paper, we discuss that the dynamic evolution of the universe
from a thermodynamic viewpoint. For this purpose, we introduce the
asymptotically holographic equipartition proposed by Padmanabhan to describe
the evolution law of the universe. In addition, we introduce the entropic
force to describe the influence of the horizon on the dynamic evolution
of the universe because there is a holographic relationship between
the bulk term and the surface term of the gravity action. Then, we
assume the equation of state of cosmic matter based on some observational
facts. Consequently, we find that the current model can well explain
the age of the universe, the relation between the luminosity distance
and the redshift factor and the current accelerated expansion
without introducing dark energy. But we should also note that the
current model introduces the ``dark energy'' when $t>t_{0}$ because
the pressure of matter is negative. We find also that the horizon
expands to the maximum when $t=2t_{0}$ and shrinks to the initial
state when $t\rightarrow4t_{0}$, so galaxies will disappear beyond
the horizon and the universe becomes increasingly dark with time progresses when $t>2t_{0}$.
By analyzing the scale factor $a(t)$, we find that there is a Big
Rip singularity.

Next, we analyze that the thermodynamic properties of current model.
We find that the generalized second law of thermodynamics always holds.
In addition, we find that the energy balance relation
also holds, which in turn shows that there is a holographic relationship
between the bulk degrees of freedom and the surface degrees of freedom
of the spacetime. We also show that the energy equipartition relation
for these degrees of freedom on the surface always holds.
Finally, we analyze the energy condition and point out
that the strong energy condition is always violated and the
weak energy condition holds when $t\leq2t_{0}$.
These results of thermodynamic analyses also confirm
that the current model can describe the dynamic behaviors and thermodynamic properties of
the universe well when $t<t_{0}$.

At the end of this paper, let us talk about the dynamic evolution of the universe by
introducing fluid viscosity or particle creation. Both models avoid
introducing dark energy to explain the accelerated expansion of the
universe. In the viscous fluid model \cite{key-59,key-60,key-61,key-62},
there is an additional pressure due to the viscosity of the fluid,
which leads to the accelerated expansion of the universe. In addition,
the viscous matter model can be mapped into the phantom dark energy
scenario with constraints under certain conditions. Furthermore, the
entropy of the universe filled with viscous fluid increases, so the
second law of thermodynamics holds. Similarly, the particle creation
model \cite{key-63,key-64,key-65,key-66,key-67} also introduces an
additional negative pressure and can describe the same dynamic and
thermodynamic behavior of the universe as $\Lambda(t)CDM$ model under
certain conditions. But compared with $\Lambda(t)CDM$ model, the
particle creation model can unify dark matter and dark energy with
one dark component. Besides, it is also shown that the entropic force model
with a matter creation rate proportional to $H$ is statistically
highly favored \cite{key-67}. Compared with these models,
we introduce the entropic force to describe the accelerated expansion of the universe.
On this basis, we build a cosmological model which can well describe the thermodynamic
and dynamic behaviors of the universe compatible with astronomical
observations based on the description of the microcosmic degrees
of freedom of the spacetime.

\section*{Acknowledgments}
This research was funded by Major Research
Project for Innovative Group of Education Department of Guizhou Province (Grant No. KY[2018]028) and the
NNSF of China (Grants No. 11775187, No. 11865018 and No. 11865019).

\end{document}